\documentclass[twocolumn,notitlepage,prl,superscriptaddress,longbibliography]{revtex4-2}
\usepackage{amsfonts}
\usepackage{textcomp}
\usepackage{times}
\usepackage{graphicx}
\usepackage{float}
\usepackage{latexsym,amsmath,amssymb,bm,euscript}
\usepackage{color}
\usepackage{subfigure}
\usepackage{epstopdf}
\usepackage[colorlinks=true,linkcolor=blue,citecolor=blue]{hyperref}
\usepackage{hyperref}
\usepackage{soul}
\usepackage[normalem]{ulem}
\usepackage{mathrsfs}
\usepackage{amsmath}
\usepackage{xspace}
\usepackage{natbib}
\usepackage{ulem}
\usepackage{etoolbox}
\usepackage{tikz}

\newcommand{\LNO}{La$_3$Ni$_2$O$_7$}

\begin{document}

\title{Bilayer $t$-$J$-$J_\perp$ Model and Magnetically Mediated Pairing in the Pressurized Nickelate La$_3$Ni$_2$O$_7$}

\author{Xing-Zhou Qu}
\thanks{These authors contributed equally to this work.}
\affiliation{Kavli Institute for Theoretical Sciences, University of Chinese
    Academy of Sciences, Beijing 100190, China}
\affiliation{CAS Key Laboratory of Theoretical Physics, Institute of
    Theoretical Physics, Chinese Academy of Sciences, Beijing 100190, China}

\author{Dai-Wei Qu}
\thanks{These authors contributed equally to this work.}
\affiliation{Kavli Institute for Theoretical Sciences, University of Chinese
    Academy of Sciences, Beijing 100190, China}
\affiliation{CAS Key Laboratory of Theoretical Physics, Institute of Theoretical Physics, Chinese Academy of Sciences, Beijing 100190, China}

\author{Jialin Chen}
\thanks{These authors contributed equally to this work.}
\affiliation{CAS Key Laboratory of Theoretical Physics, Institute of Theoretical Physics, Chinese Academy of Sciences, Beijing 100190, China}
\affiliation{Hefei National Laboratory, Hefei 230088, China}

\author{Congjun Wu}
\affiliation{New Cornerstone Science Laboratory, Department of Physics, 
School of Science, Westlake University, 310024, Hangzhou, China}
\affiliation{Institute for Theoretical Sciences, Westlake University, 310024, 
Hangzhou, China}
\affiliation{Key Laboratory for Quantum Materials of Zhejiang Province,
School of Science, Westlake University, Hangzhou 310024, Zhejiang, China}
\affiliation{Institute of Natural Sciences, Westlake Institute for Advanced Study, 310024, Hangzhou, China}

\author{Fan Yang}
\affiliation{School of Physics, Beijing Institute of Technology, Beijing 100081, China}

\author{Wei Li}
\email{w.li@itp.ac.cn}
\affiliation{CAS Key Laboratory of Theoretical Physics, Institute of
    Theoretical Physics, Chinese Academy of Sciences, Beijing 100190, China}
\affiliation{Hefei National Laboratory, Hefei 230088, China}
\affiliation{CAS Center for Excellence in Topological Quantum Computation,
    University of Chinese Academy of Sciences, Beijing 100190, China}

\author{Gang Su}
\email{gsu@ucas.ac.cn}
\affiliation{Kavli Institute for Theoretical Sciences, University of Chinese
    Academy of Sciences, Beijing 100190, China}
\affiliation{CAS Center for Excellence in Topological Quantum Computation,
    University of Chinese Academy of Sciences, Beijing 100190, China}

\begin{abstract}
The recently discovered nickelate superconductor La$_3$Ni$_2$O$_7$ 
has a high transition temperature near 80~K under pressure, providing 
an additional avenue for exploring unconventional superconductivity. 
Here with state-of-the-art tensor-network methods, we study a bilayer 
$t$-$J$-$J_\perp$ model for La$_3$Ni$_2$O$_7$ and find a robust 
$s$-wave superconductive (SC) order mediated by interlayer magnetic 
couplings. Large-scale density matrix renormalization group calculations 
find algebraic pairing correlations with Luttinger parameter $K_{\rm SC} 
\lesssim 1$. Infinite projected entangled-pair state method obtains a 
nonzero SC order directly in the thermodynamic limit, and estimates 
a strong pairing strength $\bar{\Delta}_z \sim \mathcal{O}(0.1)$. 
Tangent-space tensor renormalization group simulations elucidate the 
temperature evolution of SC pairing and further determine a high SC 
temperature $T_c^*/J \sim \mathcal{O}(0.1)$. Because of the intriguing
orbital selective behaviors and strong Hund's rule coupling in the compound, 
$t$-$J$-$J_\perp$ model has strong interlayer spin exchange (while 
negligible interlayer hopping), which greatly enhances the SC pairing 
in the bilayer system. Such a magnetically mediated pairing has also 
been observed recently in the optical lattice of ultracold atoms. Our 
accurate and comprehensive tensor-network calculations reveal a robust 
SC order in the bilayer $t$-$J$-$J_\perp$ model and shed light on the 
pairing mechanism of the high-$T_c$ nickelate superconductor.
\end{abstract}

\maketitle

\textit{Introduction.---}
High-$T_c$ superconductivity, since its discovery in doped 
cuprates~\cite{Cuprate1986,LNW-HighTc-RMP2006,Keimer2015Nature},
has raised long-lasting research interests. Very recently, under 
a high pressure of above 14 GPa, a Ruddlesden-Popper bilayer 
perovskite La$_3$Ni$_2$O$_7$ exhibits a high $T_c$ near 80~K
\cite{Nickelate80K}. Later on, optical measurements show that the 
compound features strong electronic correlations that place it in 
the proximity of a Mott phase~\cite{Liu2023correlation}, despite 
certain density-wave-like order under ambient pressure. Zero 
resistance and strange metal behaviors have been reported under 
high pressure by other experimental groups~\cite{Hou2023emergence,
Zhang2023hightemperature}.
Currently, the electronic structure, effective model, and pairing 
mechanism in the pressurized nickelate \LNO~are under very
active investigation~\cite{Luo2023bilayer,Zhang2023electronic,
Yang2023s-wave,Kuroki2023bilayer,Gu2023pairing,Werner2023correlated,
Cao2023flat,Wu2023charge,Shen2023effective,Lu2023interlayer}.

% ======= Fig. 1 ====== %
\begin{figure}[!tbp]
\includegraphics[width=1\linewidth]{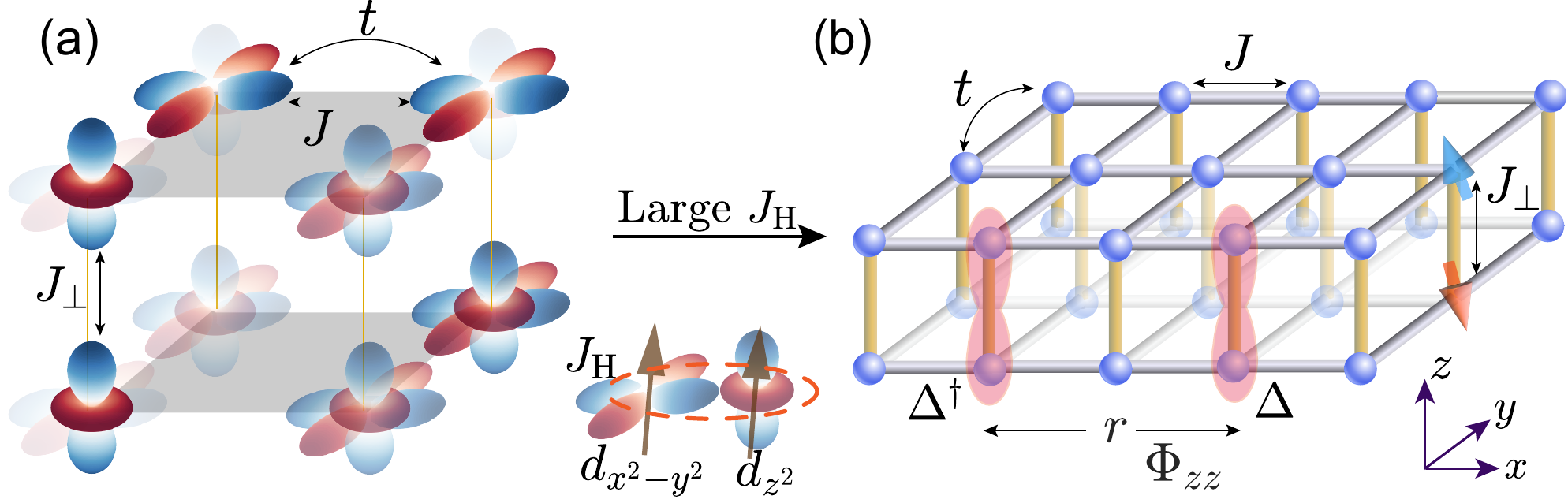}
\caption{(a) Two $e_g$ orbitals in the bilayer structure of the nickelate \LNO.
The quarter-filled $d_{x^2-y^2}$ orbitals form an effective $t$-$J$ model 
with intralayer hopping $t$ and spin exchange $J$. The $d_{z^2}$ orbital 
is localized and has an interlayer AF exchange through the $\sigma$ 
bonding. The spins of $d_{x^2-y^2}$ and $d_{z^2}$ orbitals are coupled 
through an on-site FM Hund's rule coupling $J_{\mathrm{H}}$. 
In the large $J_{\rm H}$ limit, we arrive at (b) bilayer $t$-$J$-$J_\perp$ 
model, where the interlayer AF coupling $J_\perp$ is strong while the 
interlayer hopping is absent. The SC pairing correlation $\Phi_{zz}(r)$ 
is between two interlayer pairing $\Delta^{(\dagger)}$ along the vertical 
$z$ direction and separated by a distance $r$ along the $x$ direction.}
\label{Fig1}
\end{figure}

A bilayer two-orbital Hubbard model has been proposed to 
describe the high-pressure phase of La$_3$Ni$_2$O$_7$, 
where the kinetic part is determined from the density functional 
theory calculations~\cite{Luo2023bilayer}, and the interactions
including the Hubbard $U$, Hund's rule coupling $J_{\rm H}$, etc., 
can be included. The SC instability and related pairing symmetry 
have been discussed with weak (to intermediate) coupling approaches
\cite{Yang2023s-wave,Kuroki2023bilayer,Gu2023pairing}. 
Nevertheless, the large Coulomb interaction $U/t \gg 1$ in 
\LNO~urgently calls for analysis from a strong coupling approach
\cite{Werner2023correlated,Cao2023flat,Wu2023charge}. 
Precision many-body calculations are required to scrutinize 
the possible SC order in the effective model~\cite{Shen2023effective}. 
The tensor-network methods constitute a powerful and versatile 
approach for both ground-state~\cite{White1992,Schollwock2011MPS,
Verstraete2004renorm,Cirac2021RMP,Corboz2010Simulation,
Jordan2008Classical} and finite-temperature properties
\cite{Li2011a,Chen2018XTRG,tanTRG2023,Chen2021SLU,
Chen2022tbg} of correlated electrons. Unfortunately, 
the original bilayer two-orbital model poses great challenges 
to tensor-network calculations and a dimension reduction in 
local Hilbert space while retaining the essence of electron 
correlations in the nickelate is very necessary. 

Lately it is proposed that by considering the orbital selective behaviors
of localized $d_{z^2}$ and itinerant $d_{x^2-y^2}$ electrons, together
with the strong ferromagnetic (FM) Hund's rule coupling, a bilayer 
$t$-$J$-$J_\perp$ model with strong antiferromagnetic (AF) interlayer 
exchange $J_\perp$ may provide an adequate effective model for 
\LNO~\cite{Lu2023interlayer}. Here we perform high-precision ground-state 
and finite-temperature tensor-network calculations of this bilayer model, 
and reveal a robust SC order with high $T_c$ that may account for the 
observation in the pressurized nickelate La$_3$Ni$_2$O$_7$.

\textit{Bilayer $t$-$J$-$J_\perp$ model.---}
As shown in Fig.~\ref{Fig1}, we note the two $e_g$ orbitals in \LNO, 
namely, $d_{x^2-y^2}$ and $d_{z^2}$, have distinct and orbital 
selective behaviors~\cite{Zhang2023electronic}. The $d_{z^2}$ orbital is 
almost localized with flat band structure promoted by the 
strong Hund's couplings~\cite{Cao2023flat}. Considering that $d_{z^2}$ 
orbital is only slightly doped (nearly half-filled)~\cite{Werner2023correlated,
Cao2023flat,Wu2023charge}, we can freeze their charge fluctuations 
and regard the $d_{z^2}$ electrons as local moments~\cite{Cao2023flat}.
The interlayer $\sigma$ bonding through the apical oxygen
\cite{Nickelate80K} renders a prominent interlayer AF coupling 
between the $d_{z^2}$ moments (also dubbed as the ``hidden 
dimer''~\cite{Zhang2023electronic}). On the other hand, the 
$d_{x^2-y^2}$ orbital is quarter-filled and adequately described 
by a $t$-$J$ model within each layer~\cite{Yang2023s-wave,
Gu2023pairing,Wu2023charge}. The $d_{x^2-y^2}$ orbital has 
negligible interlayer single-particle tunneling~\cite{Luo2023bilayer,
Zhang2023electronic}. However, the strong FM Hund's coupling
can bind the two $e_g$ orbitals and ``passes'' the strong interlayer 
AF coupling to the $d_{x^2-y^2}$ orbital~\cite{Lu2023interlayer}, as 
illustrated in Fig.~\ref{Fig1}(a).

To see that, we start with the model $H = H_{t-J} + H_{\rm AF}
+ H_{\rm Hund},$ where $H_{t-J}$ is the intralayer $t$-$J$ model 
of $d_{x^2-y^2}$ electrons, and $H_{\rm AF}$ denotes the AF
exchange $H_{\rm AF} = J_{\perp} \sum_i \bold{S}^{d}_{i,\mu=1}
\cdot  \bold{S}^d_{i,\mu=-1}$ between the two layers. The index
$\mu=\pm1$ labels the upper(lower) layer, and $\bold{S}^d$ 
denotes the localized $d_{z^2}$ moment. $H_{\rm Hund} = 
-J_{\rm H} \sum_{i,\mu} \bold{S}^{c}_{i,\mu} \cdot \bold{S}^{d}_{i,\mu}$
is the on-site Hund's coupling, with $\bold{S}^c$ the spin of $d_{x^2-y^2}$ 
electron. To further simplify the two-orbital model, it is noted that 
the density functional theory calculations suggest $t\simeq 0.5$~eV ($d_{x^2-y^2}$), 
$t_\perp^z \simeq 0.64$~eV ($d_{z^2}$)~\cite{Luo2023bilayer,
Zhang2023electronic}, placing the nickelate in the strong coupling 
regime by taking Hubbard $U\simeq 5$~eV (i.e., $U/t\sim10$)
\cite{Cao2023flat,Werner2023correlated}. As an intra-atomic exchange, 
the FM Hund's rule coupling is {about $J_{\rm H} \sim 1$~eV
\cite{Cao2023flat,Werner2023correlated}, clearly greater} than the 
spin exchanges $J\sim 4t^2/U \simeq 0.2$~eV and $J_\perp \simeq 
0.32$~eV, which is sufficiently strong to transfer the AF couplings 
between the two $e_g$ orbitals~\cite{Qu2023Orb}.
It is therefore sensible to take the large $J_{\rm H}$ limit and 
symmetrize the spins $\bold{S}^{d}_{i,\mu}$ and $\bold{S}^{c}_{i,\mu}$ 
of the two orbitals. The AF interlayer coupling between $d_{z^2}$ 
moments can be effectively expressed as $\bold{S}^{c}_{i,\mu=1} 
\cdot \bold{S}^c_{i,\mu=-1}$ in the symmetrized spin-triplet space. 
With this, an effective single-band bilayer $t$-$J$-$J_\perp$ model 
can be obtained~\cite{Lu2023interlayer}
\begin{eqnarray}
    H_{\rm bilayer} & = & -t
    \sum_{\langle i,j \rangle, \mu, \sigma} (c^\dagger_{i, \mu, \sigma}
    c_{j, \mu, \sigma} + H.c.) \notag \\
    & + &
    J \sum_{\langle i,j \rangle, \mu} (\bold{S}^c_{i,\mu} \cdot
    \bold{S}^c_{j,\mu} - \frac{1}{4} n_{i,\mu} n_{j,\mu}) \notag \\
    & + &
    J_{\perp} \sum_{i} \bold{S}^c_{i,\mu=1} \cdot
    \bold{S}^c_{i,\mu=-1},
    \label{Eq:b-t-J}
\end{eqnarray}
where $\sigma=\{\uparrow,\downarrow\}$ is the spin orientation,
and the vector operator $\bold{S}^c_{i,\mu} = \frac {1}{2} \,
c^\dagger_{i,\mu,\sigma} \, (\bm{\sigma}_{\sigma,\sigma'})
\, c_{i,\mu,\sigma'}$ denotes the spin of the itinerant $d_{x^2-y^2}$
electron with the Pauli matrices $\bm{\sigma}  = \{\sigma_x,
\sigma_y, \sigma_z\}$. Note the double occupancy is projected
out in the $t$-$J$-$J_\perp$ model as usual.

Below we consider the intralayer hopping $t=3$ and spin exchange
$J=1$ (taken as energy scale henceforth), and the interlayer AF
couplings $J_\perp$ is varied to explore the  SC and possibly competing 
change density wave (CDW) orders. Interlayer hopping $t_{\perp}$ 
is forbidden [except in Fig.~\ref{Fig3}(c)], different from the previously 
studied bilayer Hubbard-like models~\cite{Scalapino2011bilayer}. 
As the $d_{x^2-y^2}$ orbitals are nearly quarter-filled, we set 
$n_e=0.5$ and the hole density $n_h =1 -n_e = 0.5$ in the 
pristine \LNO.

% ======= Fig. 2 ====== %
\begin{figure}[!tbp]
\includegraphics[width=1\linewidth]{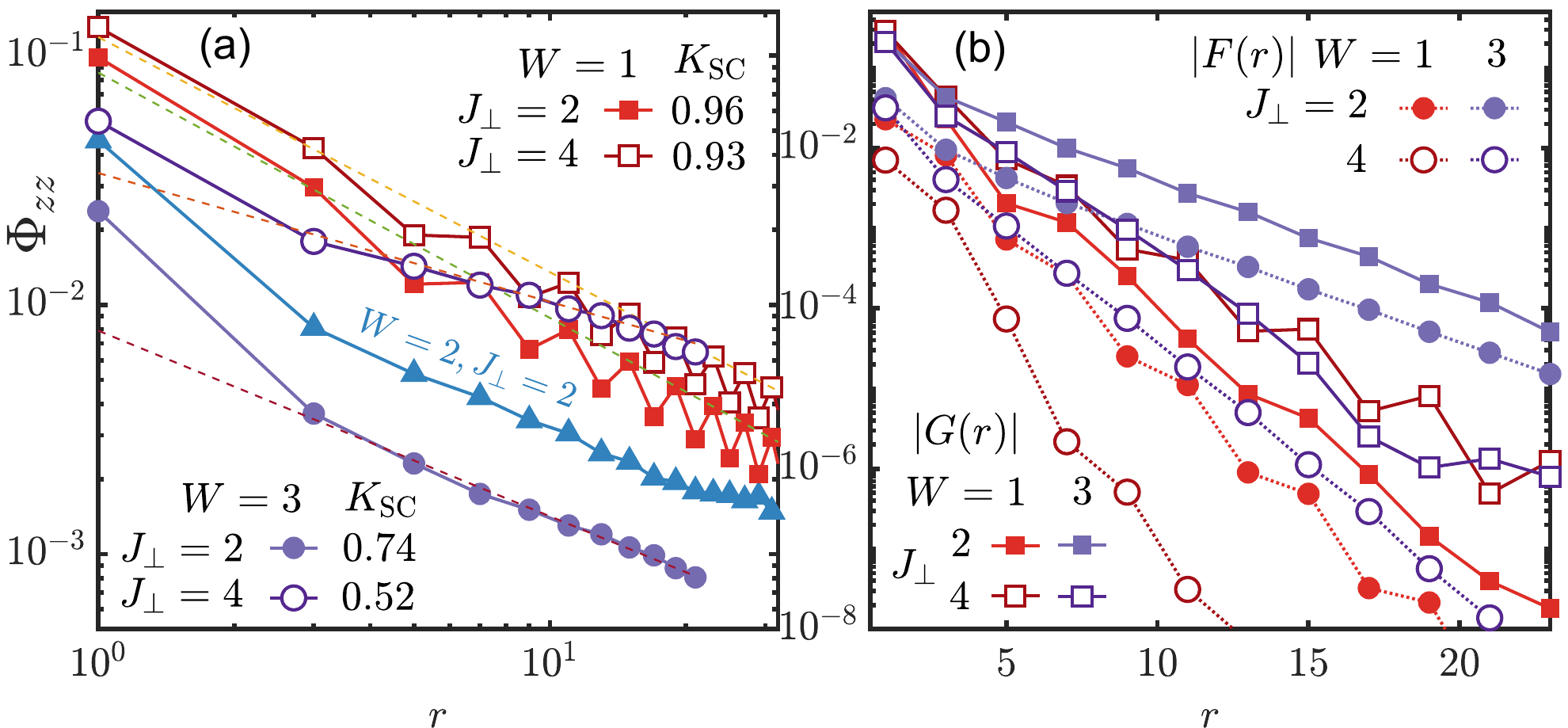}
\caption{(a) Pairing correlation $\Phi_{zz}$ on the $2\times W \times L$ 
bilayer lattices with widths $1 \leq W \leq 3$ and long length $L$, namely, 
$2 \times 1 \times 128$ ($W=1, n_e = 0.5$), $2 \times 2 \times 64$
($W=2, n_e \approx 0.54$), and $2 \times 3 \times 48$ ($W=3, 
n_e=0.5$). The SC correlations exhibit algebraic behaviors as 
$\Phi_{zz}(r)\sim r^{-K_{\rm SC}}$, enhanced with interlayer coupling 
$J_\perp$. The $W=2$ data fall into algebraic scaling with 
oscillations~\cite{SM}, leading to inaccurate extraction of the 
Luttinger parameters. (b) Spin-spin correlation $F(r)$ and the 
single-particle Green's function $G(r)$ decay exponentially 
(see definitions in the main text) in the SC phase~\cite{FN}.
}
\label{Fig2}
\end{figure}

\textit{Tensor-network methods for zero- and finite-temperature
properties.---} To simulate the bilayer model in Eq.~(\ref{Eq:b-t-J}), 
we employ tensor-network approaches for both $T=0$ and $T>0$ 
calculations. Regarding the ground state, we exploit the density matrix 
renormalization group (DMRG)~\cite{White1992,Schollwock2011MPS} 
for the finite-size systems and the infinite projected entangled-pair state 
(iPEPS) directly in the thermodynamic limit~\cite{Verstraete2004renorm,
Cirac2021RMP,Corboz2010Simulation,Jordan2008Classical}. 
In DMRG we map the $2\times W \times L$ bilayer system into a 
quasi-1D chain with long-range interactions~\cite{SM}, and implement 
the non-Abelian and Abelian symmetries with tensor libraries
\cite{Weichselbaum2012,Weichselbaum2020,ITensor,ITensor-r0.3}. 
We retain up to $D^*=12\,000$ U(1)$_{\rm charge} \times$ 
SU(2)$_{\rm spin}$ multiplets (equivalently $D\simeq 30\,000$ 
individual states), which well converge the results~\cite{SM}.
For iPEPS calculations, we adopt the simple update
\cite{Xiang2008SU,Li2012SU} with retained bond dimension 
up to $D=12$, which is extrapolated to infinite $D$ and 
compared to the DMRG results. Moreover, we exploit the 
finite-$T$ tensor networks, in particular the recently developed
tangent-space tensor renormalization group~\cite{tanTRG2023} 
to study the bilayer system with $W=1$ and length up to $L=128$. 
Up to $D^*=1600$ U(1)$_{\rm charge} \times$ SU(2)$_{\rm spin}$ 
multiplets (equivalently $D\simeq 3600$ states) render very well 
converged results down to a low temperature $T/J \simeq 0.1$~\cite{SM}.

\textit{Robust SC order and magnetically mediated interlayer pairing.---}
In Fig.~\ref{Fig2} we show the DMRG results of pairing correlations
$\Phi_{zz}(r) = \langle \Delta_{i}^\dagger \Delta_{j} \rangle$ with
interlayer pairing $\Delta_i^{\dagger} = \frac{1}{\sqrt{2}} \sum_{\mu=\pm1} 
\, c_{i,\mu,\uparrow}^{\dagger} c_{i,-\mu,\downarrow}^{\dagger}$ 
and distance $r\equiv |j-i|$, where we find $\Phi_{zz}(r)$ shows 
algebraic scaling with the Luttinger exponent $K_{\rm SC} \lesssim 1$ 
for moderate to strong $J_\perp$. In Fig.~\ref{Fig2}(b), 
we calculate the spin-spin correlation
$F(r) = \frac{1}{2} \sum_\mu \langle  \bold{S}^c_{i,\mu}
    \cdot \bold{S}^c_{j,\mu} \rangle$ and the Green's function
$G(r) = \frac{1}{4} \sum_{\mu, \sigma} \langle c^\dagger_{i,\mu,\sigma}
    c_{j,\mu,\sigma} + \mathrm{H.c.} \rangle$,
and find both correlations decay exponentially. The DMRG results 
in Fig.~\ref{Fig2} indicate the emergence of Luther-Emery liquid
\cite{Luther1974} with quasi-long-range SC order, 
as well as ﬁnite spin and single-particle gaps.

% ======= Fig. 3 ====== %
\begin{figure}[!tbp]
\includegraphics[width=1\linewidth]{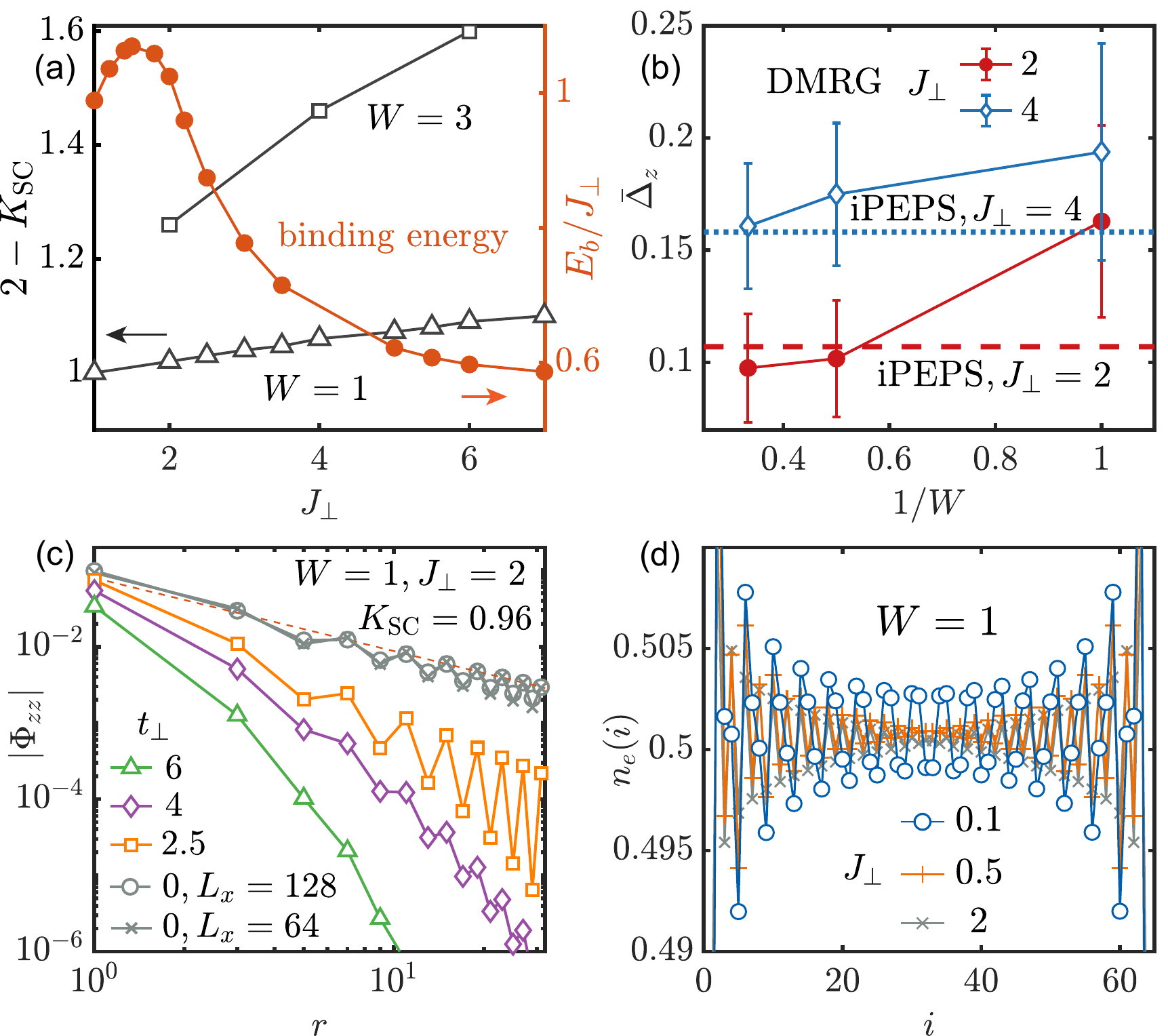}
\caption{(a) The Luttinger parameter $(2-K_{\mathrm{SC}})$ vs. $J_{\perp}$ 
calculated on various bilayer systems. The binding energy for $W=1$ system 
is also shown. (b) The SC order parameter $\bar{\Delta}_z$ obtained with 
iPEPS (dashed horizontal lines) and DMRG (solid lines) for $J_{\perp} = 2,4$. 
In the latter, $\bar{\Delta}_z$ is estimated within central columns, and 
the error bars represent the difference between the maximal and minimal 
values by varying the number of columns involved. We set $n_e=0.5$ for 
the $W=1,3$ cases, while for the $W=2$ case it is shifted slightly to 
$n_e\simeq0.54$.
(c) Pairing correlation $|\Phi_{zz}|$ with various interlayer hopping $t_\perp$.
For the $t_\perp=0$ case, we show the pairing correlations in excellent 
data convergence as computed with $L_x=64$ and 128.
(d) The electron density profiles $n_e(i), 1 \leq i \leq L$ for various 
$J_\perp$ computed on the $2 \times 1 \times 64$ system. 
}
\label{Fig3}
\end{figure}

The AF exchange $J_{\perp}$ plays an essential role in mediating 
the interlayer pairing and forming the rung-singlet SC phase
\cite{Wu2017DiagramQMC}. In Fig.~\ref{Fig3}(a), we provide 
the DMRG results of the Luttinger parameter $K_{\rm SC}$ 
controlling the scaling behaviors of pairing correlations. We find 
$(2 - K_{\mathrm{SC}}) \gtrsim 1$ increases rapidly with $J_\perp$ 
and signifies a diverging susceptibility at low temperature as 
$\chi_{\rm SC} \sim 1/T^{(2-K_{\mathrm{SC}})}$. The pairing 
susceptibility $\chi_{\rm SC} = \frac{2}{N} \partial \langle \Delta_{\rm tot} \rangle_\beta /\partial h_p$ 
measures the response of SC order parameter to a small pairing 
field $h_p$ coupled to $\Delta_{\rm tot} = \frac{1}{2} \sum_i (\Delta_{i} 
+ \Delta_{i}^\dagger)$. To further characterize the enhancement of 
pairing strength, we compute the binding energy 
$E_b = E(N_e+1) + E(N_e-1) - 2E(N_e)$, where $E(N_e)$ is the 
ground-state energy with $N_e$ electrons. In Fig.~\ref{Fig3}(a), 
we find $E_b$ increases with $J_{\perp}$ as the ratio $E_b/J_\perp \gtrsim 0.6$. 
However, $E_b/J_\perp$ is not monotonic and has a round peak at 
$J_{\perp} \approx 1.5$. In the strong $J_\perp$ limit the escalation 
of binding energy slows down its pace and the ratio converges to 
$E_b/J_\perp \simeq 0.6$ .

With iPEPS calculations directly in the thermodynamic limit where 
symmetry breaking is allowed to occur, we evaluate the SC order parameter 
$\bar{\Delta}_z = \langle \Delta^{(\dagger)}_i \rangle$ averaged 
over the two sublattices, and show the results in Fig.~\ref{Fig3}(b). 
We find $\bar{\Delta}_z$ increases with $J_\perp$ and reaches 
about 0.11 for $J_\perp =2$ and 0.16 for $J_\perp =4$. In Fig.
\ref{Fig3}(b) we also show the DMRG estimation of the order parameter 
$\bar{\Delta}_z = \sqrt{\frac{1}{N_b} \sum_{i,j} \langle \Delta_{i}^\dagger 
\, \Delta_{j} \rangle}$ where $i, j$ are restricted within $N_c$ 
central columns, and $N_b$ is number of the credited pairs. 
In practice, we vary $N_c$ from 8 to 16 for different lattice geometries, 
and find the DMRG and iPEPS results agree well. 
Notice that the order parameter $\bar{\Delta}_z$ of the 
bilayer $t$-$J$-$J_\perp$ system [$\mathcal{O}(10^{-1}$)] is much 
greater than that found in the plain $t$-$J$ square lattice 
[$\mathcal{O}(10^{-2}$)]~\cite{Corboz2014Competing}.

The order parameter $\bar{\Delta}_z$ and pairing correlations are 
found to be uniform in each layer, i.e., it belongs to an $s$-wave SC 
order. We have also computed the intralayer pairings $\Phi_{yy}$ and 
$\Phi_{yx}$ with DMRG, and the order parameters $\bar{\Delta}_{x,y}$ 
with iPEPS~\cite{SM}, which are found to be negligibly small when 
compared to $\Phi_{zz}$ (and $\bar{\Delta}_z$). Based on the results 
in Figs.~\ref{Fig2} and \ref{Fig3}, we conclude there exists a robust 
rung-singlet SC order mediated and enhanced by magnetic couplings 
$J_\perp$ in the bilayer $t$-$J$-$J_\perp$ model for \LNO.

\textit{Pauli blocking and charge density-wave instability.---}
The existence of strong interlayer $J_\perp$ while absence 
of hopping $t_\perp$ is a key for the robust SC order
to appear in the bilayer system. In Fig.~\ref{Fig3}(c) we 
artificially introduce the interlayer hopping $t_\perp$, 
and find the SC order gets weakened and even suppressed
as $t_\perp$ increases.
This can be ascribed to the Pauli blocking effect where the holes tend 
to repel each other kinetically according to their hopping amplitude
\cite{Hilker2023pairing}, spoiling the interlayer pairing for strong $t_\perp$. 
Moreover, this observation may also be relevant to the experiments: 
further increasing pressure in the SC phase of \LNO~does not enhance 
$T_c$ but decreases it~\cite{Nickelate80K,Hou2023emergence,
Zhang2023hightemperature}. It is possible that high pressure enhances 
interlayer tunneling of the $d_{x^2-y^2}$ orbitals and thus weakens 
the SC order.

In Fig.~\ref{Fig3}(d), we show the electron density distribution $n_e(i)$
by tuning $J_\perp$ to smaller values. For $J_\perp=2$, the CDW
fluctuation is rather weak, consistent with a robust $s$-wave SC state. 
However, for smaller $J_\perp$ the SC order becomes 
weakened, while the CDW instability turns strong. This may explain 
the absence of SC order in \LNO~under ambient pressure, where 
certain density-wave-like instability bas been observed in recent 
experiments~\cite{Liu2023correlation,Hou2023emergence,
Zhang2023hightemperature}. The change of interlayer Ni-O-Ni 
bond angle [from 168$^\circ$ (ambient) to 180$^\circ$ (pressurized)] 
and length (by 1.9 \AA) may sensitively influence $J_\perp$, 
thus switching between the CDW and SC phases. Moreover, 
by reducing the hole density we find even clearer CDW pattern
\cite{SM}, suggesting that the CDW instability or stripe phase 
may also be a competing order in the bilayer nickelate.

% ======= Fig. 4 ====== %
\begin{figure}[!tbp]
\includegraphics[width=1\linewidth]{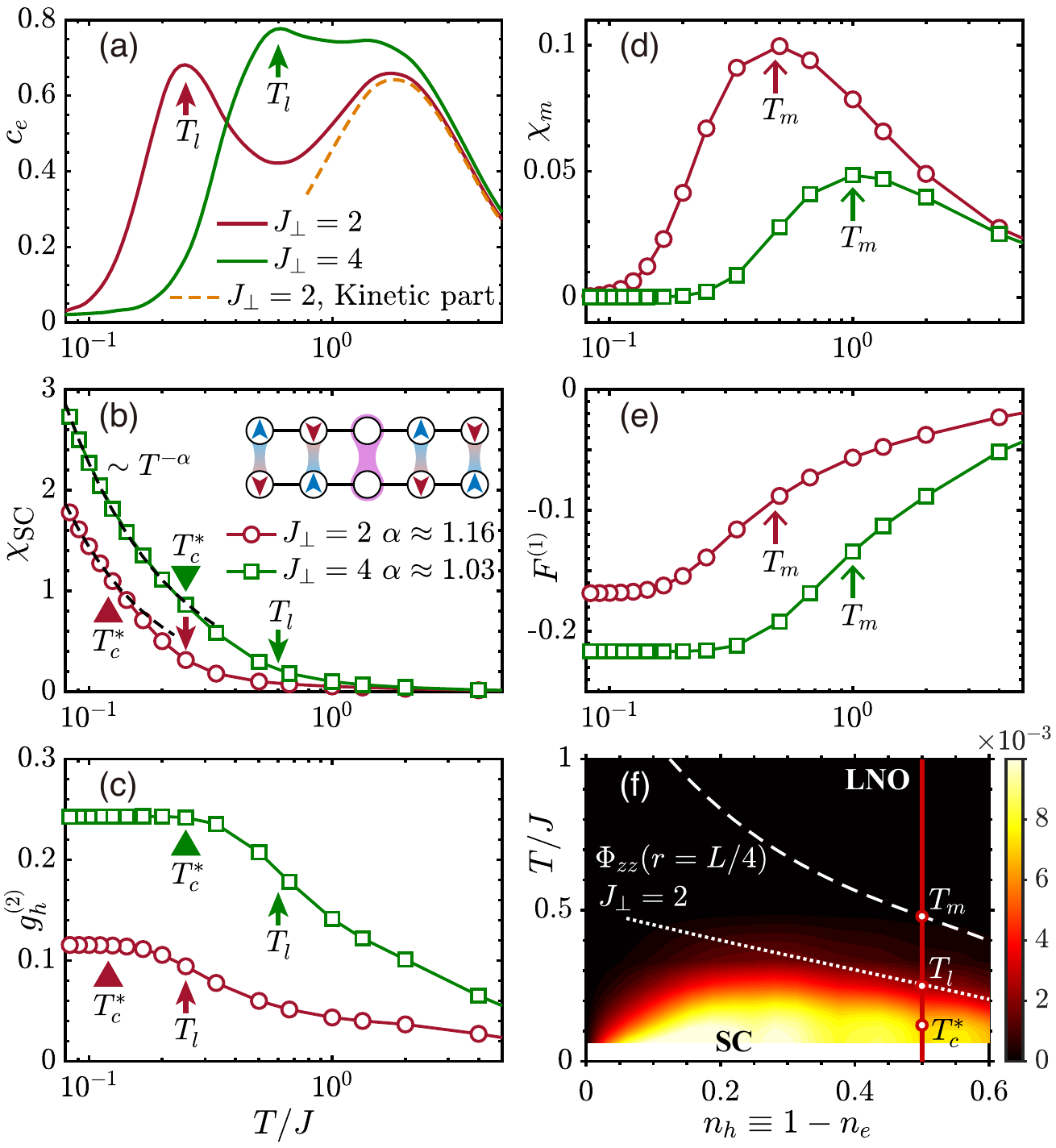}
\caption{Finite-temperature results for $W=1$ systems with
length up to $L=128$. In panels (a)-(e), the data for $J_\perp = 2$ 
(red lines) and $4$ (green) are shown. In (a),(c)-(e) the hole 
density is fixed as $n_h \equiv 1-n_e = 0.5$, and in (b) the 
data are calculated with fine-tuned chemical potential that 
leads to $n_h \simeq 0.5$. (a) shows the specific heat $c_e$ 
with $T_l$ the lower characteristic temperature. 
(b) shows the pairing susceptibility $\chi_{\rm SC}$, which 
diverges with a power-law scaling $T^{-\alpha}$ (the dashed 
lines) for $T \leq T_c^*$. The interlayer pairing 
and AF correlations are illustrated in the inset.
(c) shows the interlayer hole correlations on the rungs, with 
$T_l$ and $T_c^*$ determined from (a) and (b), respectively. 
(d) shows the magnetic susceptibility $\chi_m$ with a hump 
at $T_m$. (e) shows the rung spin correlations. 
(f) The contour plot of the pairing correlation $\Phi_{zz}(r=L/4)$ 
for various hole densities $n_h$ computed with $J_\perp = 2$. 
The vertical red line denotes the $n_h=0.5$ case relevant for 
\LNO~(LNO). 
}
\label{Fig4}
\end{figure}

\textit{Finite-temperature pairing and magnetic susceptibilities.---}
In Fig.~\ref{Fig4} we show the temperature evolutions of spin and 
pairing correlations. Firstly, from Fig.~\ref{Fig4}(a) we find the electron 
specific heat $c_e = \frac{1}{N_e} \frac{\partial \varepsilon}{\partial T}$ 
exhibits a double-peak structure, with the higher-$T$ peak contributed 
by the kinetic energy, and lower-$T$ scale responsible for SC pairing 
labeled by $T_l$. In Fig.~\ref{Fig4}(b) we apply a uniform pairing field 
$-h_p \Delta_{\rm tot}$ with $h_p = 2 \times 10^{-3}$, and 
compute the pairing susceptibility. We find that $\chi_{\rm SC}$ is 
rather small for $T > T_l$ and becomes significant for $T < T_l$, 
{making $T_l$} the SC-fluctuation onset temperature.

As temperature further lowers, we find $\chi_{\rm SC}$ exhibits an 
algebraic divergence $\chi_{\rm SC} \sim T^{-\alpha}$ for temperature 
below $T_c^*$, when the system enters the low-temperature SC 
regime. The fitted exponent in Fig.~\ref{Fig4}(b) is $\alpha\simeq1$, 
consistent with the ground-state DMRG results of $K_{\rm SC}\simeq 1$ 
for $W=1$. Both $T_l$ and $T_c^*$ increase with $J_\perp$,
and in Fig.~\ref{Fig4}(a) we find $T_l/J \simeq 0.25$ for 
$J_\perp=2$, which is enhanced to $T_l/J \simeq 0.6$ for 
$J_\perp=4$. Similarly in Fig.~\ref{Fig4}(b) the $\chi_{\rm SC}$ 
curves show an overall enhancement, and the $T_c^* /J$ increases 
from $0.12$ for $J_\perp=2$ to about $0.25$ for $J_\perp=4$. 
In Fig.~\ref{Fig4}(c) we show the hole-hole correlation 
$g_h^{(2)} =(2/N)\sum_i \langle h_{i,\mu=1} h_{i,\mu=-1} \rangle_\beta/ (\langle h_{i,\mu=1}
\rangle_\beta \cdot \langle h_{i,\mu=-1} \rangle_\beta) - 1$, 
where $N$ is the number of lattice sites with 
$h_{i,\mu}$ the hole density operator. The positive $g_h^{(2)}$ 
indicates the attractive (``bunching'') correlations between the holes. 
From Fig.~\ref{Fig4}(c), we find $g_h^{(2)}$ rapidly increases 
at about $T_l$ and saturates at about $T_c^*$ when the pairing 
susceptibility starts to diverge algebraically in Fig.~\ref{Fig4}(b).

To further reveal the intriguing interplay between antiferromagnetism 
and superconductivity in the bilayer system, we compute the magnetic 
susceptibility $\chi_m$ and rung spin-spin correlation $F^{(1)} = \frac{2}{N} 
\sum_i \langle \bold{S}_{i, \mu=1} \cdot \bold{S}_{i, \mu=-1} \rangle_\beta$ 
in Fig.~\ref{Fig4}(d),(e). The magnetic susceptibility $\chi_m$ becomes
suppressed below $T_m$ in Fig.~\ref{Fig4}(d),  which can be ascribed
to the rapid establishment of correlation $F^{(1)}$ at about the same 
temperature [Fig.~\ref{Fig4}(e)].

\textit{Temperature evolution of the SC order.---}
Now we summarize the temperature evolution of the pairing correlations 
in Fig.~\ref{Fig4}(f), where the red line denotes \LNO~with hole density
$n_h \equiv 1- n_e \simeq 0.5$. As temperature lowers, the interlayer AF 
correlation develops at about $T_m / J \simeq 0.48$, and then the hole 
bunching occurs at $T_l / J \simeq 0.25 $, shortly after that the system 
enters the coherent regime below $T_c^* /J \simeq 0.12$, establishing 
eventually the quasi-long-range SC order.

As shown in Fig.~\ref{Fig4}(f), by doping electrons into the system (or 
via a self-doping from $d_{z^2}$ to the $d_{x^2-y^2}$ orbitals), the SC 
order and its characteristic temperature can be further enhanced. 
For $J_\perp=2$, the optimal hole density appears at $n_h \sim 
0.25$, to the electron-doping side of \LNO, as also evidenced 
by the enhanced pairing susceptibility and temperature scales $T_l$ 
and $T_c^*$~\cite{SM}.

\textit{Discussion and outlook.---}
We exploit multiple tensor-network methods and reveal robust $s$-wave
SC order in the bilayer $t$-$J$-$J_\perp$ model for the recently discovered 
nickelate superconductor.
In \LNO~the $d_{x^2-y^2}$ orbital has a hole density of $n_h\simeq0.5$
---a large value on the verge of quenching the SC in cuprates. 
For the latter, large hole doping may undermine the intralayer AF correlations 
and suppress the SC order. Surprisingly, in \LNO~ the SC order remains 
robust and has a high $T_c\simeq 80$~K even with large hole density. 
{Based on our $t$-$J$-$J_\perp$ model calculations, we ascribe it to 
the robust pairing mechanism mediated by strong interlayer AF exchange. 
In Fig.~\ref{Fig4}(f), we find indeed the SC dome can extend to a 
very wide regime up to $n_h \sim 0.6$ for the bilayer nickelate.}

We would also point out an intriguing and rather unexpected 
connection between the high-$T_c$ nickelate and ultracold atom 
systems. Recently, {the mixed dimensional (mixD) bilayer optical 
lattices with strong interlayer spin exchange while no interlayer 
single-particle tunneling has been realized~\cite{Grusdt2022mixD,
Hilker2023pairing}.} Remarkably, such a $t$-$J$-$J_\perp$ mixD 
bilayer model naturally emerges in the orbital-selective nickelate \LNO: 
The $d_{x^2-y^2}$ electrons are itinerant within each layer, 
while $d_{z^2}$ orbitals are nearly half-filled and localized. 
The FM Hund's coupling ``glues'' the spins of two $e_g$ orbitals, 
conveying to the $d_{x^2-y^2}$ electrons a strong AF coupling---
the driving force for the interlayer pairing.  
To thoroughly validate our effective model for \LNO, a detailed 
analysis of the two-orbital bilayer model with realistic parameters 
is necessary. Our preliminary results support the scenario proposed
here~\cite{Qu2023Orb}.

Overall, our results provide a solid and valuable basis 
for understanding the unconventional SC in pressurized \LNO~from 
a strong coupling approach, and put various experimental observations 
in a coherent picture. They offer useful guidance for future studies in 
the nickelate superconductors and also mixD ultracold atom systems.

\begin{acknowledgments}
W.L. and F.Y. are indebted to 
Yang Qi and Qiaoyi Li for stimulating 
discussions. This work was supported by the National 
Natural Science Foundation of China (Grants No. 12222412, 
No. 11834014, No. 11974036, No. 12047503, No. 12074031, No. 12174317, and 
No. 12234016), Strategic Priority Research Program of CAS 
(Grant No.~XDB28000000), Innovation Program for Quantum 
Science and Technology (No. 2021ZD0301800 and No. 2021ZD0301900), 
the New Cornerstone Science Foundation, and CAS Project for 
Young Scientists in Basic Research (Grant No.~YSBR-057). 
We thank the HPC-ITP for the technical support and generous 
allocation of CPU time.
\end{acknowledgments}

\bibliography{tJRefs}

\begin{thebibliography}{48}
\bibitem [{FN()}]{FN}%
  \BibitemOpen
  \href@noop {} {\bibinfo {title} {{We compute the spin, charge, and SC pairing
  correlations on a pair of sites separated by distance $r$ and symmetric
  around the center of the system to avoid boundary effects.}}}\BibitemShut
  {Stop}%
\end{thebibliography}

%======================================
%========== Supplementary =============
%======================================
\newpage
\clearpage
\onecolumngrid
\mbox{}
\begin{center}

   {\large Supplemental Materials for} $\,$ \\
   \bigskip
   \textbf{\large{Bilayer $t$-$J$-$J_\perp$ Model and Magnetically Mediated
           Pairing in the Pressurized Nickelate La$_3$Ni$_2$O$_7$}} \\

   Qu \textit{et al}.
\end{center}

\date{\today}

\setcounter{section}{0}
\setcounter{figure}{0}
\setcounter{equation}{0}
\renewcommand{\theequation}{S\arabic{equation}}
\renewcommand{\thefigure}{S\arabic{figure}}
\setcounter{secnumdepth}{3}

\section{Zero and Finite-temperature Simulation Methods}
In this work, we employ the density matrix renormalization group (DMRG)
and infinite projected entangled-pair state (iPEPS) for the ground state,
as well as the tangent-space tensor renormalization group ($\tan$TRG) for
finite-temperature properties of the bilayer $t$-$J$-$J_\perp$ model.

\subsection{DMRG Calculations}
In the DMRG calculations, two different tensor-network libraries have
been employed. We exploit QSpace library~\cite{Weichselbaum2012,
    Weichselbaum2020} to implement $\mathrm{U(1)}_{\text{charge}} \times \,
    \mathrm{SU(2)}_{\text{spin}}$ symmetry in our DMRG code. ITensors
library~\cite{ITensor,ITensor-r0.3} is also utilized to perform a
$\mathrm{U(1)}_{\text{charge}} \times \mathrm{U(1)}_{\text{spin}}$
DMRG as a double check. For example, in the $2 \times 1 \times 128$
systems we keep up to 2000 U(1)$_{\rm charge}$ $\times$
    SU(2)$_{\rm spin}$ multiplets, which ensures a small truncation error
$\lesssim 10^{-11}$. In the $2 \times 3 \times 48$ systems
    we keep up to 12 000 U(1)$_{\rm charge}$ $\times$ SU(2)$_{\rm spin}$
    multiplets with typical truncation error $\sim 5 \times 10^{-6}$.
In the calculations, $D$ represents the number of U(1)$_{\rm charge}$
$\times$ U(1)$_{\rm spin}$ states and $D^*$
for the U(1)$_{\rm charge}$ $\times$ SU(2)$_{\rm spin}$ multiplets. 

% ============================== %
\begin{figure}[!htbp]
    \includegraphics[width=1\linewidth]{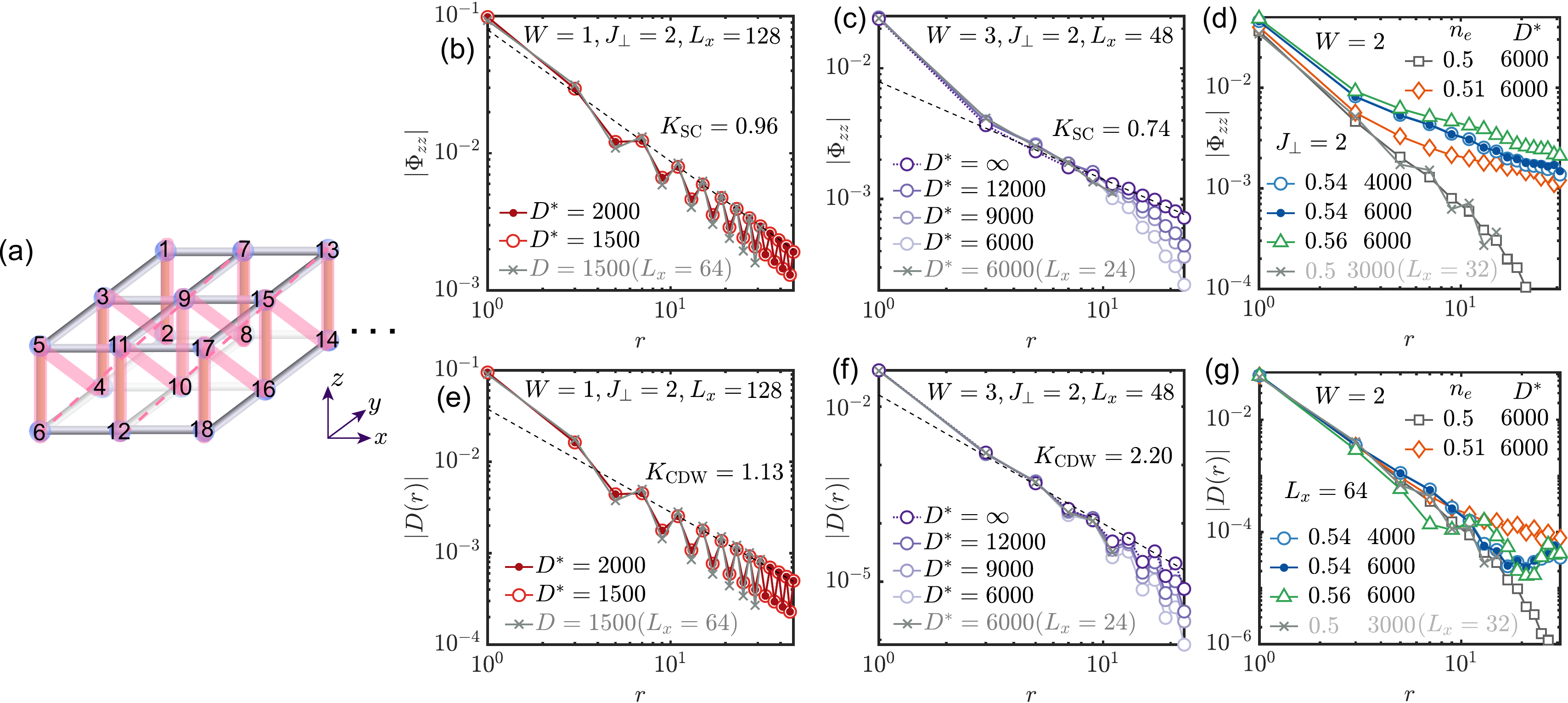}
    \caption{(a) Illustration of the zigzag path used in DMRG calculation,
        and the model parameters are chosen as $J_\perp=2$, $J = 1$, and $t = 3$.
        The DMRG results on the $t$-$J$-$J_\perp$ model on the $2\times W\times L_x$ lattices with longer $L_x$ are shown in (b-g). The (b,c,d) panels show the pairing correlations $\Phi_{zz}(r) = \langle \Delta^\dagger_i \Delta_j \rangle$ and (e,f,g) show the charge correlations $D(r) = \langle n_i n_j \rangle - \langle n_i \rangle \langle n_j \rangle $, with $r\equiv |i-j|$. We show $W=1,2,3$ systems with relatively long $L_x$ up to 128. As a comparison, the results with shorter length $L_x$ are displayed with grey lines and cross markers. Note that for $W=1,3$ cases, the Luttinger parameters $K_{\rm SC}$ and $K_{\rm CDW}$ can be accurately extracted. On the other hand, though the $W=2$ data fall into algebraic decay, they are highly oscillating and lead to inaccurate extractions of the Luttinger parameters.
    }
    \label{FigS1}
\end{figure}

\begin{figure}
    \includegraphics[width=0.9\linewidth]{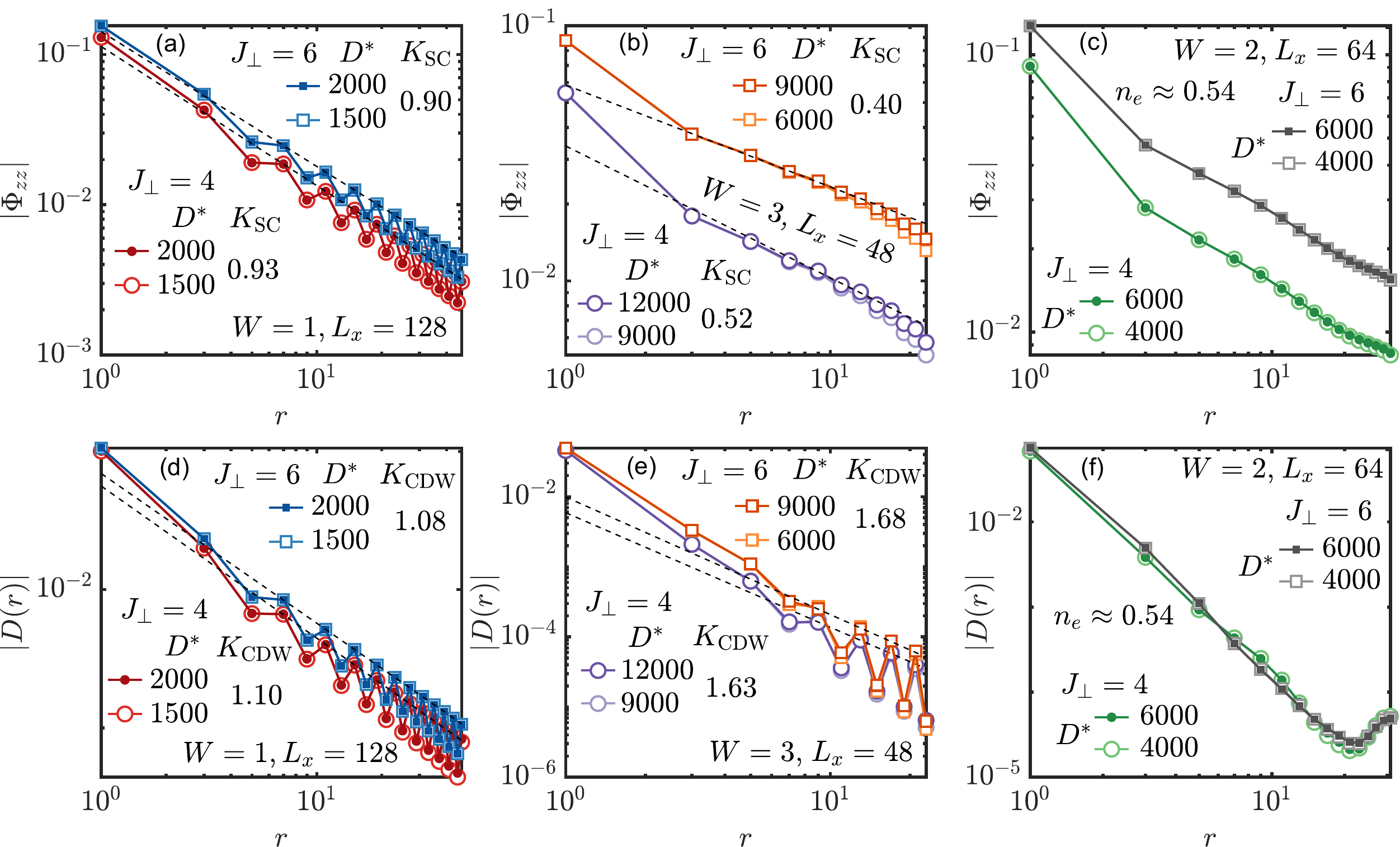}
    \caption{DMRG results of the $W=1, 2, 3$ system with two different interlayer
            couplings $J_\perp = 4, 6$. (a, d) pairing correlation and charge correlation for the
            $W=1, n_e =0.5$ system, (b, e) for $W=3, n_e = 0.5$ system and (c, f) for $W=2,
                n_e \approx 0.54$ system.}
    \label{FigS2}
\end{figure}

\textit{Data convergence.---}
In Fig.~\ref{FigS1} we compare the results obtained in different system
sizes and bond dimensions to confirm the convergence of our ground-state
simulations. The pairing and charge correlations are presented in
    Fig.~\ref{FigS1} for $J_\perp = 2$ and in Fig.~\ref{FigS2} for $J_\perp = 4, 6$
    systems. For the $W=1$ case [Fig.~\ref{FigS1}(b,e) and Fig.~\ref{FigS2}(a, d)],
we find the data are fully converged by retaining $D^*=2000$ (equivalently
$D \approx 4400$ individual states), from which we can extract the Luttinger
parameters $K_{\rm SC}$ and $K_{\rm CDW}$ accurately. For the $W=3$
case shown in Fig.~\ref{FigS1}(c,f), we retain up to 12 000 bond multiplets,
and find that the correlations are converged for short distances. For longer
distances, we perform a fitting with polynomial function $C(1/D^*) = C(0) +
    a/D^* + b/{D^*}^2$ and extrapolate the data to infinite $D^*$. With this we
obtain an accurate estimation of $K_{\rm SC}$ and $K_{\rm CDW}$ indicated
in Fig.~\ref{FigS1}(c,f). Since the DMRG simulations converge faster on
    $J_\perp = 4, 6$ systems, we did not extrapolate the data and rely on the well converged
    segment to extract the Luttinger parameters.

\textit{$W=2$ case.---}
    The results of $W=2$ case are shown in Figs.~\ref{FigS1}(d,g)
    (for $J_\perp=2$) and \ref{FigS2}(c,f) (for $J_\perp=4,6$). Firstly,
        we point out that there has a subtlety at exact quarter-filling (i.e.,
    $n_e=0.5$). The pair correlations computed on the $L_x =32$ system
    turn out to decay exponentially when extending the calculations to longer
    systems [grey lines in Fig.~\ref{FigS1}(d,g)].  An explanation for this
    phenomenon could be that the system is being viewed as two coupled
    $t$-$J$ ladders with a finite charge gap at quarter filling, as reported in
    Ref.~\cite{Lu2022tJ}. Therefore, we fine tune the electron density
    to $n_e \simeq 0.51$-$0.56$ for the $W=2$ case in Figs.~\ref{FigS1}(d,g)
    and \ref{FigS2}(c,f), in order to avoid the peculiar filling $n_e=0.5$ for
    the $W=2$ geometry. In Figs.~\ref{FigS1}(d) and \ref{FigS2}(c), we now
    observe robust superconductivity for the $W=2$ case. It is worth noting
    that the chosen $n_e \gtrsim 0.5$ is actually more closely related to the
    realistic electron density in the $d_{x^2-y^2}$ orbital due to the self-doping
    \cite{Werner2023correlated,Lu2023interlayer}, and it also leads to more
    consistent results for different widths $W$ [as can be seen in Fig.~3(b)
            in the main text].

\textit{CDW correlations.---}
    In Fig.~\ref{FigS1}(e,f,g) and Fig.~\ref{FigS2}(d,e,f) we also show
    results of the charge correlations. For the $W=1$ case ($J_\perp=2$),
    we find in Fig.~\ref{FigS1}(e) that the density correlation $D(r) = \langle
        n(i) n(i+r) \rangle - \langle n(i) \rangle \langle n(i+r) \rangle$ exhibits a
    power-law scaling with Luttinger parameter $K_{\rm CDW} \simeq 1.13$.
    Moreover, we find the product of two Luttinger parameters $K_{\rm SC}
        \, K_{\rm CDW} = 1.08 \approx 1$, which fulfills the expectation of the LE
    theory. Likewise, the $W=1, J_\perp = 4, 6$ cases shown
        in Fig.~\ref{FigS2}(a, d) also meet $K_{\rm SC} \, K_{\rm CDW} \approx 1$.
    However, for wider systems, such as the $W=3, J_\perp = 2$ case
    shown in Fig.~\ref{FigS1}(f) and $W=3, J_\perp = 4, 6$ case in
    Fig.~\ref{FigS2}(e) , we find that the extracted $K_{\rm CDW}$
    is much larger than the SC parameter $K_{\rm SC}$. This indicates
    a uniform SC order with relatively weak CDW correlations in wider
    systems, suggestive of a robust SC order in the 2D limit. Regarding
    the $W=2$ case, either for $J_\perp=2$
    [Fig.~\ref{FigS1}(g)] or $J_\perp=4,6$ [Fig.~\ref{FigS2}(f)], we see
    strong oscillations and quite difficult to extract the Luttinger exponent
    $K_{\rm CDW}$ there.

% ============================== %
\begin{figure}[!]
    \includegraphics[width=0.8\linewidth]{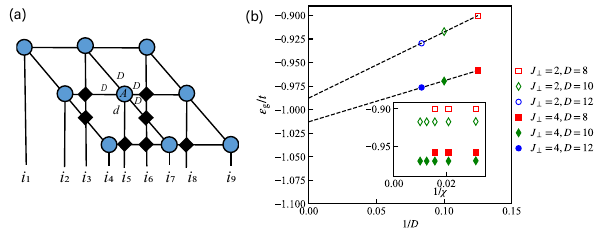}
    \caption{(a) Illustration of fermionic iPEPS ansatz used in our calculations,
        where the local tensors (blue solid circles) and the swap gates (black
        diamonds) taking care of the fermion sign are indicated. $D$ and $d$
        represent the bond dimensions of the geometric and physical indices.
        (b) Energy per site $\varepsilon_\mathrm{g}/t$ obtained by iPEPS as a
        function of  the inverse bond dimension $1/D$, the environment bond
        dimension is chosen as $\chi=D^2$. The inset shows the convergence
        of $\varepsilon_\mathrm{g}/t$ as a function of $1/\chi$ for various bond
        dimension $D$. Other model parameters are $J = 1$ and $t = 3$.}
    \label{iPEPS_Eg}
\end{figure}

\subsection{iPEPS method}
To simulate the $t$-$J$-$J_\perp$ model directly in the thermodynamic
limit, we flatten the bilayer system into a single layer system with
enlarged local Hilbert space and use the fermionic iPEPS method
\cite{Corboz2010Simulation,Corboz2009Fermionic,Barthel2009Contraction,
    Kraus2010Fermionic}. The wavefunction ansatz is illustrated in
Fig.~\ref{iPEPS_Eg}(a), with a $2\times 2$ unit cell consisting of
two bulk tensors (see inset of Fig.~\ref{iPEPS_Delta} below). Each
bulk tensor has four virtual bonds whose dimension $D$ can control
the simulation accuracy (up to $D=12$ in practice). It also has a physical
bond representing $d=9$ local electron configurations on the flattened
single-layer system. We optimize the iPEPS wavefunction using
simple update~\cite{Xiang2008SU,Li2012SU,Corboz2010Simulation}
and the expectation values are calculated using corner transfer
matrix renormalization group method~\cite{Corboz2014Competing,
    Orus2009Simulation} with environment bond dimension $\chi=D^2$
that leads to converged results as shown in the inset of Fig.~\ref{iPEPS_Eg}(b).
It also shows that the energy per site $\varepsilon_\mathrm{g}/t$
approaches infinite-$D$ limit linearly with $1/D$ for both $J_\perp=2$
and $J_\perp=4$ cases.

% ============================== %
\begin{figure}[!tbp]
    \includegraphics[width=0.75\linewidth]{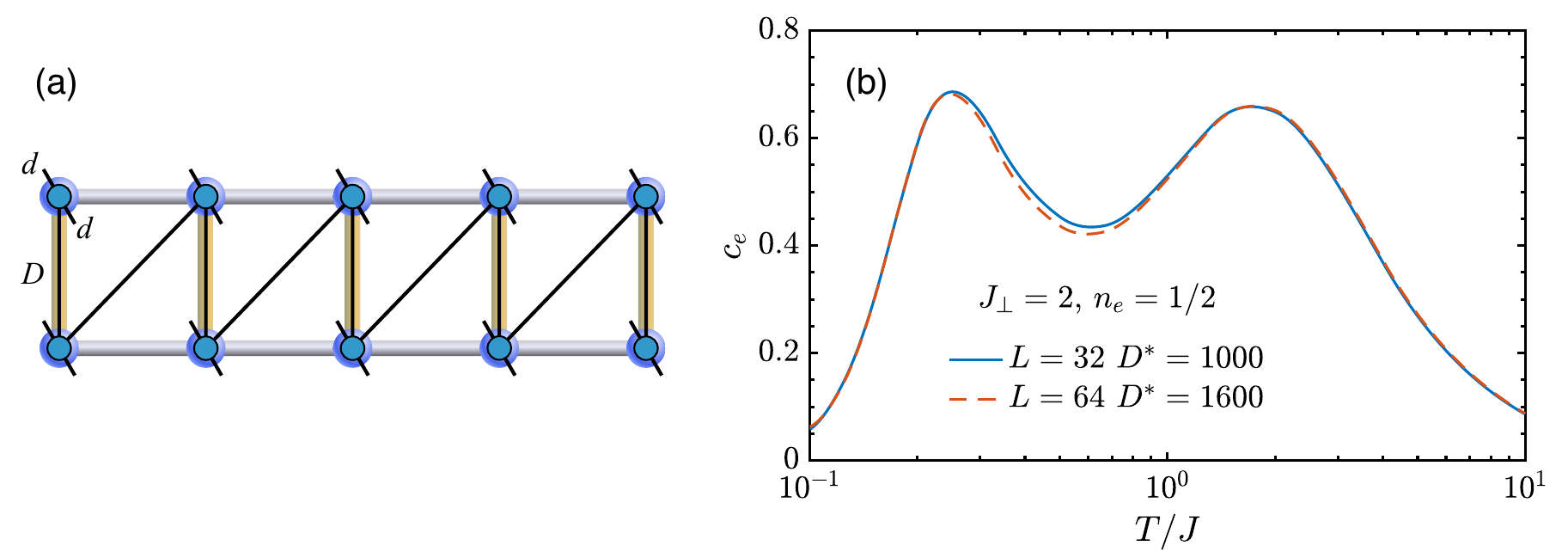}
    \caption{(a) In $\tan$TRG calculations, we map the $2\times1\times L$
        system to a quasi-1D one along the zigzag path. The thermal density
        operator is represented as an MPO, which consists of rank-4 tensors
        with two geometric indices with bond dimension $D$ and two physical
        indices with bond dimension $d=3$.
        (b) Electron specific heat $c_e$ for the $2\times1\times L$ systems with
        $J_\perp=2$ and electron density $n_e=1/2$. Results computed with
        different system sizes and bond dimensions are compared, i.e., $L=32$
        and $D^*=1000$ (multiples, equivalent $D\simeq 2200$ individual states)
        vs $L=64$ and $D^*=1600$ ($D\simeq 3600$). We find they agree with
        each other, showing very good convergence over the system sizes and
        bond dimensions.}
    \label{FigS:Convergence}
\end{figure}

\subsection{$\tan$TRG method}
To simulate the finite-temperature properties of the bilayer systems,
we use the tangent-space tensor renormalization group ($\tan$TRG)
approach~\cite{tanTRG2023}, which is a state-of-the-art finite-$T$
approach for many-electron problems. In this approach, the system
is mapped to quasi-one-dimensional geometry, and the thermal
density operator is represented as a matrix product operator (MPO),
as shown in Fig.~\ref{FigS:Convergence}(a). In practical calculations,
the Abelian and non-Abelian symmetries are implemented with the
QSpace library~\cite{Weichselbaum2012,Weichselbaum2020}. To
compute the specific heat, magnetic susceptibility, and various electron
correlations, we implement $\mathrm{U(1)}_{\text{charge}} \times
    \mathrm{SU(2)}_{\text{spin}}$ symmetry and keep up to $D^*=1600$
multiplets on systems up to $L=64$. To compute the pairing susceptibility
$\chi_\textrm{SC}$, we have added a small pairing field, which breaks
the particle number conservation. Therefore, we use
$\mathbb{Z}_{2, \textrm{charge}} \times \mathrm{SU(2)}_\textrm{spin}$
symmetry and retain up to $D^*=1000$ multiplets, on long system up to
$L=128$. As the grand canonical ensemble is used in our thermal tensor
network simulations, a chemical potential term $-\mu \sum_i n_i$ is added
to fine tune the electron density $n_e$.

To check the convergence of our thermal data, in Fig.~\ref{FigS:Convergence}(b)
we show the electron specific heat $c_e$ results for the $2\times1\times L$ system
with $J_\perp=2$ and electron density $n_e=1/2$, computed on two different system
sizes and bond dimensions: $L=32$ with $D^*=1000$, and $L=64$ with $D^*=1600$.
We find the $c_e$ changes little for the two cases and see a very good convergence
of the thermal data.

\begin{figure}[!htbp]
    \includegraphics[width=0.6\linewidth]{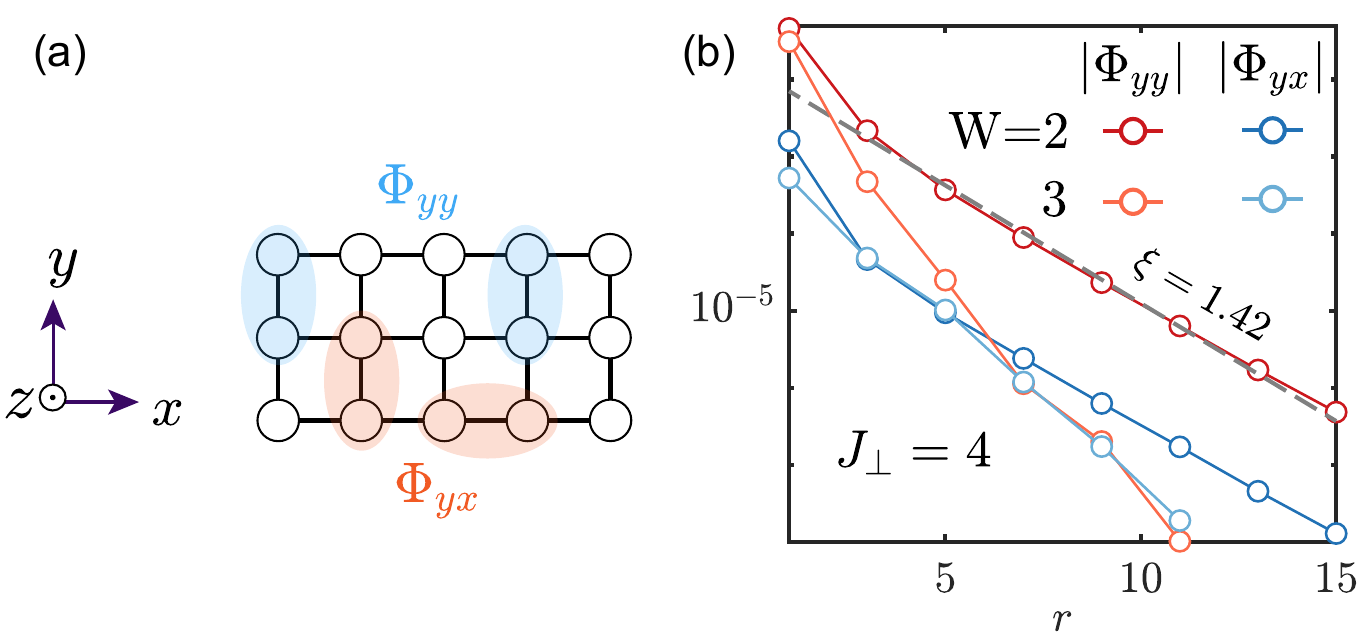}
    \caption{
        (a) Schematic of the intralayer pairing correlations $|\Phi_{yy}|$ and $|\Phi_{yx}|$.
        (b) Intralayer pairing correlations
        calculated in $2\times 2 \times 32$ and $2\times 3\times 24$ systems,
        from which we can estimate the correlation length $\xi \simeq 1.42$
        for $|\Phi_{yy}|$ on $W=2$ lattice. Other lines have similar or even
        shorter correlation lengths.
    }
    \label{DMRG_interlayer}
\end{figure}

% ======= Supplementary Numerical Results ======= %
\section{More Numerical Simulation Results}
Below we provide supplemental results obtained with DMRG ($T=0$, finite system),
iPEPS ($T=0$, thermodynamic limit), and the $\tan$TRG ($T>0$) calculations.

\subsection{DMRG results}
We show in Fig.~\ref{DMRG_interlayer} the intralayer pairing correlation
$\Phi_{yy}$ and $\Phi_{yx}$ on the $2\times 2 \times 32$ and
$2\times 3\times 24$ systems. The exponentially decaying
pairing correlations indicate the absence of intralayer pairing.

\begin{figure}[!htbp]
    \includegraphics[width=0.65\linewidth]{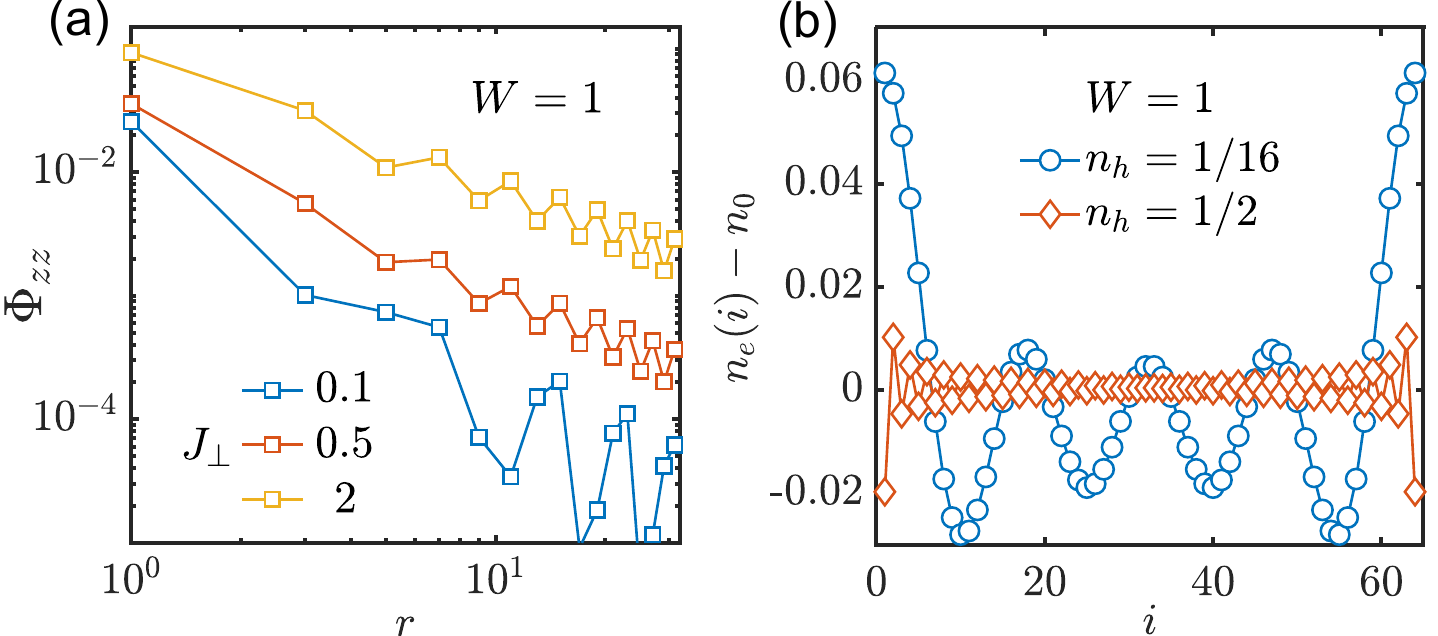}
    \caption{(a) The pairing correlations $\Phi_{zz}$ are shown for different
        $J_\perp = 0.1, 0.5$ and $2$.
        (b) The CDW profiles for different hole densities $n_h=1/2$ and $1/16$.}
    \label{DMRG_CDen}
\end{figure}

In Fig.~\ref{DMRG_CDen}(a) we find the pairing correlation $\Phi_{zz}$
decreases as interlayer AF coupling $J_\perp$ weakens. The rung-single SC
order becomes very weak for $J_\perp=0.1$, supporting our conclusion that the
interlayer pairing in the system is mediated by the interlayer AF couplings.
In Fig.~\ref{DMRG_CDen}(b) we find the charge density wave (CDW)
becomes stronger as the hole density $n_h$ decreases, which is stabilized
and even constitutes a long-range pattern for $n_h = 1/16$.

\begin{figure}[!htbp]
    \includegraphics[width=0.45\linewidth]{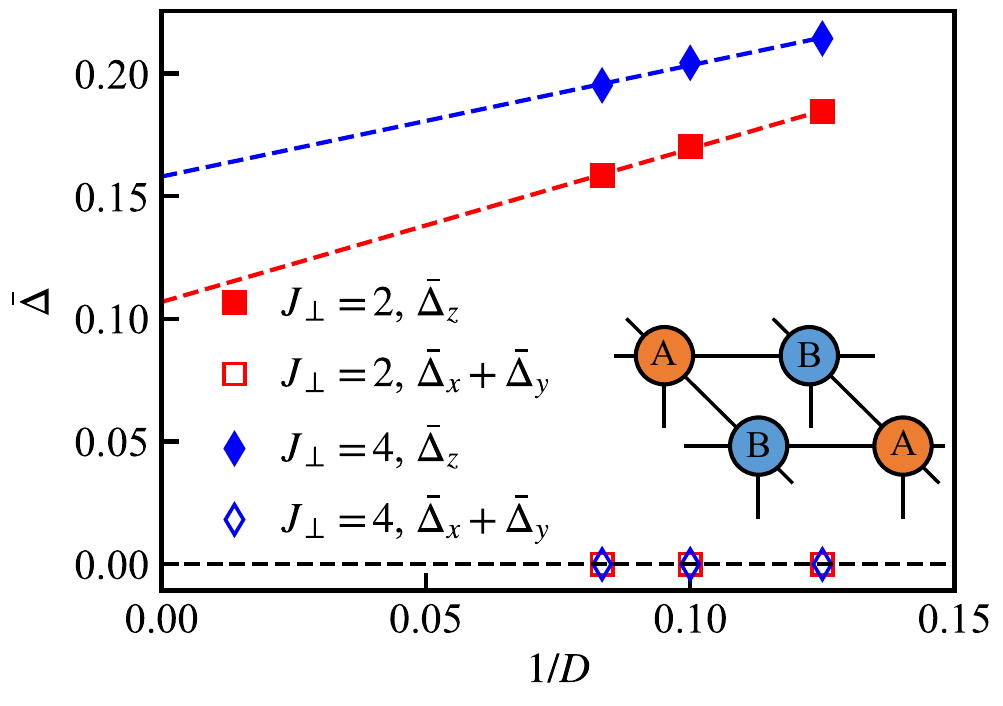}
    \caption{The rung-singlet SC order parameters obtained by iPEPS are shown
        as a function of inverse $D$. $\bar{\Delta}_{z}$ represents the interlayer
        pairing along the $\hat z$ direction, while $\bar{\Delta}_{x}$ and $\bar{\Delta}_{y}$
        represent the intralayer pairing along the $\hat x$ and $\hat y$ directions,
        respectively. The inset is an illustration of the unit cell of the iPEPS
        ansatz with two bulk tensors: A and B. Each bulk tensor has one physical
        bond and four virtual bonds.
        The $\bar{\Delta}_z$ values are computed and averaged over two bulk
        tensors, and then extrapolated to infinite $D$ limit. $\bar{\Delta}_z$ has a finite
        and rather large value, while $|\bar{\Delta}_{x(y)}|$ values are negligibly small.}
    \label{iPEPS_Delta}
\end{figure}

\subsection{iPEPS results}
We show in Fig.~\ref{iPEPS_Delta} the SC order parameter $\bar{\Delta}$
obtained by the iPEPS method in the infinite systems with different $J_\perp$.
The amplitudes of $\bar{\Delta}_{z} =\frac{1}{\sqrt{2}} \langle
    \sum_{\mu=\pm1} \, c_{i,\mu,\uparrow}^{\dagger}
    c_{i,-\mu,\downarrow}^{\dagger} \rangle$ for both
$J_\perp=2$ and $J_\perp=4$ decrease with increasing bond dimensions
$D$, while still have very large values when extrapolated to the infinite-$D$ limit.
It indicates that the interlayer pairing persists in this limit. On the other hand,
we do not find intralayer pairing as the amplitudes of $\bar{\Delta}_x$ and
$\bar{\Delta}_y$ are negligible for both $J_\perp$ values, where $\bar{\Delta}_{x(y)}
    = \frac{1}{\sqrt{2}} \sum_{\sigma=\{\uparrow, \downarrow\}} \langle
    \mathrm{sgn}{(\sigma)} \, c_{i,\mu=\pm1,\sigma}^{\dagger} c_{i+\hat{x}(\hat{y}),
            \mu=\pm1,\bar{\sigma}}^{\dagger} \rangle$, with $\rm{sgn}(\uparrow)=1$,
$\mathrm{sgn}(\downarrow)=-1$, and $\hat{x}$($\hat{y}$)
unit vector in the square-lattice plane. $\bar{\sigma}$ reverses the spin
orientation of $\sigma$. In Fig.~\ref{iPEPS_Delta} we find that
$\bar{\Delta}_z$ is enhanced by increasing $J_\perp$ from $2$ to $4$.
The iPEPS results also support the conclusion that a larger $J_\perp$
strengthens the interlayer pairing.

\subsection{$\tan$TRG results}
\begin{figure}[!tbp]
    \includegraphics[width=0.85\linewidth]{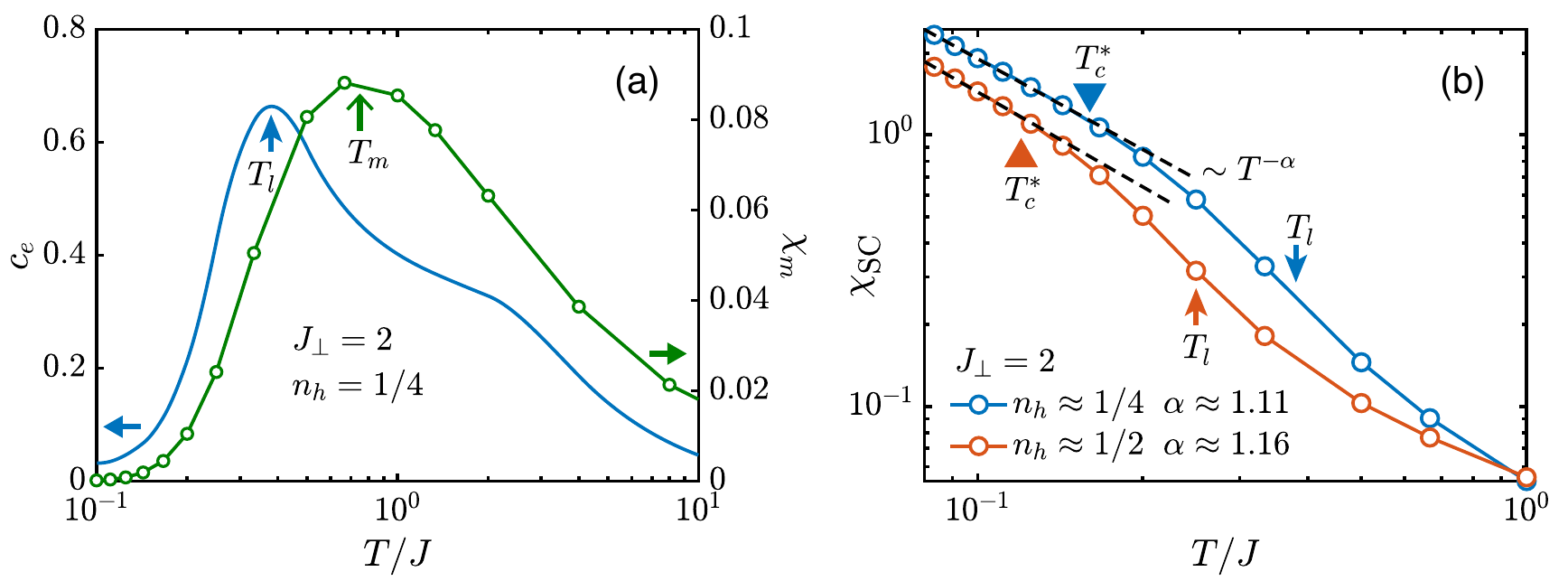}
    \caption{(a) Specific heat $c_e$ and magnetic susceptibility $\chi_m$ for the
        $2\times 1\times 32$ system with $J_\perp=2$ and hole density $n_h=1/4$.
        The blue upward arrow indicates the lower temperature scale in $c_e$.
        The green upward arrow indicates the peak temperature of $\chi_m$.
        (b) Pairing susceptibility for the $2\times 1\times 128$ system with $J_\perp=2$
        and $n_h=1/4$, $1/2$. The $n_h=1/2$ data is exactly the same one
        as in Fig.~4(b) of the main text, taken here as a comparison. The dashed
        lines denote the power-law fitting $T^{-\alpha}$. The corresponding
        temperature scales $T_c^*$ and $T_l$ are indicated.}
    \label{FigS:Ce}
\end{figure}

In Fig.~4 of the main text, we have shown the $c_e$ and $\chi_m$ for $J_\perp=2$
with hole density $n_h=1/2$. Here in Fig.~\ref{FigS:Ce}(a) we show the $c_e$
and $\chi_m$ results for the same $J_\perp=2$ but with smaller hole density
$n_h=1/4$.
Now the low-temperature peak of $c_e$ is found to be located at
$T_l/J \simeq 0.38$, and the hump of $\chi_m$ is at $T_m/J \simeq 0.75$.
Both temperature scales are higher than the corresponding values in the
quarter filling ($n_h=n_e=1/2$) case shown in Fig.~4 of the main text.

In Fig.~\ref{FigS:Ce}(b) we show the pairing susceptibility $\chi_{\rm SC}$
for $J_\perp=2$ and hole density $n_h=1/4$, and compare it to the $n_h=1/2$
case. We find there is an overall enhancement in $\chi_{\rm SC}$
when $n_h$ is lowered from $1/2$ to $1/4$. Moreover, the characteristic
temperature scale $T_c^*$ for superconductivity increases from
$T_c^*/J\simeq 0.12$ to $0.16$. Therefore, compared with quarter
filling case with $n_h=1/2$ (pristine La$_3$Ni$_2$O$_7$), the
superconductivity is enhanced by lowering hole density ($n_h=1/4$).

\end{document}